\documentclass[useAMS,usenatbib]{mn2e}
\usepackage{times}
\usepackage{graphicx}

\title[Is it possible to reveal the lost siblings of the Sun?]{Is it possible
 to reveal the lost siblings of the Sun?}
\author[Mishurov \& Acharova]
{ Yu. N. Mishurov,$^{1,2}$\thanks{E-mail:
unmishurov@sfedu.ru (YuNM); iaacharova@sfedu.ru (IAA)} 
and I.A.Acharova,$^{1}$\\
$^{1}$ Department of Physics of Cosmos, Southern Federal University, 
5 Zorge, Rostov-on-Don, 344090, Russia \\
$^{2}$ Special Astrophysical Observatory of the Russian Academy of Sciences, 
N.Arkhyz, Karachaevo-Cherkessia, 369167, Russia}

\begin{document}

\date{Accepted 2010 xxxx. Received 2010 xxxx; in original form 2010 xxxx}

\maketitle

\pagerange{\pageref{firstpage}--\pageref{lastpage}} \pubyear{2010}
       
\label{firstpage}

\begin{abstract}

We present the results of our numerical experiments on stellar scattering in the galactic disc under the influence of the perturbed galactic gravitation field connected with the spiral density waves and show that the point of view according to which stars do not migrate far from their birthplace, in general, is incorrect. Despite close initial locations and the same velocities after 4.6 Gyrs members of an open cluster are scattered over a very large part of the galactic disc. If we adopt that the parental solar cluster had $\sim 10^3$ stars, it is unlikely to reveal the solar siblings within 100 pc from the Sun. The problem stands a good chance to be solved if the cluster had $\sim 10^4$ stars. 

We also demonstrate that unbound open clusters disperse off in a short period of time under the influence of spiral gravitation field. Their stars became a part of the galactic disc. We have estimated typical times of the cluster disruption in radial and azimuth directions and the corresponding diffusion coefficients.

\end{abstract}
                     
\begin {keywords}
Galaxy: solar neighborhood - stars: kinematics and dynamics - Galaxy: structure
\end{keywords}
                     
\section {Introduction}

The fate and history of the Sun have a special significance since its influence on the Earth. Moreover, as it was demonstrated by Bland-Hawthorn \& Freeman (2004) and Bland-Hawthorn et al. (2010) in-depth study of our star and discovery its siblings (the stars which were born along with the Sun in the same open cluster) will supply us by a valuable information about the early epoch of our galaxy Milky Way formation and its evolution.

In a recent paper, Portegies Zwart (2009) has revised the problem of searching the solar family. According to Portegies Zwart, if we adopt that the parental cluster had $\sim10^3$ stars, about 10 to 60 solar siblings are now located within a distance of 100 pc,  
              \footnote {In what follows we call the region within 100 pc from a star as ``the close vicinity".}
              so they can be identified by means of next generation telescopes (see the first attempt of searching the solar relatives in Brown et al. 2010).
                    
The presence of solar parental cluster members in the close vicinity from the Sun is a necessary but not sufficient condition to reveal them. It is obvious, in order to recognize the siblings among $\sim 10^5$ close stars we need some additional marks like strong chemical elements similarity to the solar abundances, special velocity pattern in space, etc. (see the above papers).

In the modeling, Portegies Zwart (2009, see also Brown et al. 2010) adopts that both the Sun and its cluster members move along close and almost circular trajectories. Consequently, the authors derive an obvious result: the best place to find the solar kin is a ring-like segment close to the solar trajectory. Moreover, according to Portegies Zwart opinion one should not expect that the siblings can drift far from the Sun independently of their birth-place within the cluster since stellar velocity dispersion in the parental cluster is small ($\sim 1$ km $s^{-1}$).
       
The reason why Portegies Zwart and his coworkers came to their conclusion is obvious: they consider an oversimplified axisymmetric model for the galactic gravitation field. The above assumptions, introduced by Portegies Zwart and Brown et al. in their model, are broken up if we take into account stellar interactions with perturbations of the galactic gravitation field due to spiral density waves, especially near the corotation resonance. Indeed, as it was shown by Le\'pine et al. (2003) and Acharova et al. (2004), in the vicinity of the resonance, a star is subjected to a great wander over the galactic radius ($\sim$ 3 kpc). So, we expect that in the course of solar life-time, open clusters will diffuse in the galactic disc under the influence of spiral perturbations of the galactic gravitation field and, in general, they will be scatter over a very large part of the disc.

The goal of the present paper is to demonstrate the above effects.

\section {Model description}

We consider motion of an ensemble of probe particles, imitating stars of the solar parental cluster, in the galactic gravitation field perturbed by the density waves, which are responsible for spiral arms. So, the potential of the galactic gravitation field $\varphi_G$ is represented as a sum:
$$\varphi_G = \varphi_0 + \varphi_S,$$ where $\varphi_0$ is the unperturbed axisymmetric part, $\varphi_S$ is its perturbation due to galactic spiral density waves. 

Notice here. We are not intended to reconstruct the evolution of the Galaxy as in the cited above papers by Bland-Hawthorn and his co-authors. Our aim is narrower: we only plan to demonstrate that, during the solar life-time ($\sim$~4.6 Gyr), stars with close initial coordinates can be scattered over a very large part of the galactic disc due to interactions with the perturbed gravitation field so that, in general, in the close vicinity of any star from the parental cluster we do not reveal many its siblings as it was proclaimed by Portegies Zwart (2009). That is why we do not simultaneously consider the galactic evolution and stellar diffusion but suppose the perturbed potential as being given. 

In our approach the unperturbed gravitation force is in balance with the centrifugal one (the galactic disc rotates in its plane), i.e.: $d\varphi_0/dr = r\Omega^2$, $\Omega(r)$ is the angular rotation velocity of the disc, $r$ is the galactocentric distance. The perturbed gravitation potential we write in a form of running spiral wave:
$$\varphi_S = A cos(\chi).$$
Here $A < 0$ is the wave amplitude, 
$$\chi = m[cot(p) log(r/r_0) - \vartheta +\Omega_Pt]$$
is the wave phase (the condition $\chi = const$ is the equation for a spiral arm shape), $m$ is the number of spiral arms, $p$ is the pitch angle of arms (for trailing arms $p < 0$), $\vartheta$ is the azimuth angle in the galactic plane (for simplicity we only consider stellar trajectories lying in the galactic plane), $\Omega_P$ is the angular rotation velocity of density waves, $t$ is time, $r_0$ is a normalizing radius. It is important to recall, that whereas the galactic matter rotates differentially (i.e. $\Omega$ is a function of $r$), density waves rotate as a solid body ($\Omega_P = const$). The distance $r_C$ where both the velocities coincide, i.e. 

\begin{equation}
\Omega(r_C) = \Omega_P, 
\end{equation}
is called the {\it corotation} radius. Here the particles resonantly interact with spiral perturbations of the galactic gravitation field.

Several words about the details of the galactic gravitation field model. For 
the rotation curve we adopt the same model by Allen and Santill\'an (1991) 
with the scale for the present solar galactocentric distance $r_S = 8.5$ 
kpc as in Portegies Zwart and his collaborators papers (the normalizing radius, 
$r_0$, in the above expression for $\chi$, $r_0 = r_S$). 

In our paper, we consider both the {\it quasi-stationary} density waves 
(Lin et al. 1969) and {\it transient} ones (e.g. Sellwood \& Binney 2002 and papers therein). In the both cases the spiral wave amplitude, $A$, is defined by means of fixing the perturbed force amplitude 

\begin{equation}
F = mcot(p)A/(\Omega r)^2|_{r_S} = 0.1
\end{equation}
at solar distance (L\'epine et al. 2003; Acharova et al. 2004; Antoja et al. 2009). 

To determine the value of $\Omega_P$ we set the location of the corotation radius, $r_C$. After that, for the given rotation curve, $\Omega_P$ is computed by means of equation (1). For the quasi-stationary wave pattern 2 values of the corotation radius were examined:  
1)~the corotation resonance is situated close to the present solar 
position ($r_C = r_S = 8.5$ kpc, see e.g. Mishurov et al. 1997; Amaral \& L\`epine 1997; Mishurov \& Zenina 1999 a,b; L\'epine et al. 2001 and papers therein) and  
2)~$r_C = 3.5$ kpc (e.g. Weinberg 1994; Dehnen 2000).

For the number of spiral arms 3 models were considered: 
1)~pure 2-armed ($m = 2$) structure; 
2)~pure 4-armed ($m = 4$) structure and 
3)~superposition of $m = 2 +4$ wave harmonics (L\'epine et al. 2001, details see below).

If the galactic spiral structure is a succession of transient density waves we assume that the corotation radius is confined within the region from 4 to 15 kpc, correspondingly for the adopted rotation curve $\Omega_P$ belongs to the interval 14.2 -- 51.8 km s$^{-1}$ kpc$^{-1}$. Further, we generate random values of $\Omega_P$ uniformly distributed in the above region supposing that a spiral event of pattern speed $\Omega_P$ is steady during life-time $\tau_P$ for which 2 values were considered: $\tau_P = 100$ and 500 Myr. In all generated events we suppose patterns to be two-armed ($m = 2$) with amplitudes and pitch angles as above.

Now let us discuss the initial conditions for the probe particles. To formulate them Portegies Zwart (2009), first of all, integrated the backward (in time) solar motion and found out its birth-place that he adopts as the initial location for the parental cluster center. 
  \footnote {As it was said in Introduction, that could be done because Portegies Zwart adopts a very simple model for the galactic gravitation field neglecting by perturbations due to spiral density waves. It is well known that in an axisymmetric galactic field stellar trajectories in the galactic plane represent simple epicyclic lines (cycloid like), the trajectories in this case being stable in the sense that small deviations in initial coordinates of stars do not lead to a significant response in their final locations, at least in radial directions.}
Then, Portegies Zwart randomly distributed 1000 probe stars on the galactic plane within a circle of radius 3 pc centered at the above computed point.

But if we take into account the perturbations due to density waves the trajectories happen to be very tangled (especially near the corotation resonance). More over, as it will be shown below they are very sensible to the initial locations of particles. Hence we can not indicate more or less unambiguously the birth-place of the Sun despite the galactic gravitation field is regular (in the case of transient spiral structure the situation is much worse because of the random values of $\Omega_P$).

Hence, unlike Portegies Zwart we assume that at the initial moment of time the cluster center is located at some point with coordinates $(r_{cc},\vartheta_{cc})$. Further, following Portegies Zwart, we randomly distribute 1000 probe particles with constant surface density in a ring region on the galactic plane centered at the above point and of radius 3 pc. Then the stars were assigned the same rotation velocities corresponding to the galactocentric distance $r_{cc}$ plus the additional peculiar velocity -10.1 km/s in radial direction and 15.5 km/s in the direction of galactic rotation independently of their positions.
   \footnote{ In the cluster, stellar velocities are additionally randomly 
    perturbed. But the results do not significantly depend on the inclusion of 
    these perturbations since the chaotic velocities of the cluster members are 
    small.}
After that, we compute the final locations of the cluster members at present time assuming that the solar and cluster age to be 4.6 Gyrs.

Since we cannot indicate more or less exactly the solar birth-place, instead of estimation the number of siblings in the close vicinity of the Sun, we compute a number of stars from the parental cluster which have the siblings in their close vicinity and estimate the number of kin.

Simultaneously our experiments enable to quantitatively estimate the rate of stellar diffusion in radial and azimuth directions after the open cluster started to fall apart.

       \section {Results and discussion}
       
We performed a lot of experiments trying to understand the scatter of stars under the influence of perturbations from spiral gravitation field associated with galactic density waves. Below will be presented some of our model results derived for various parameters. However, to compare in relief our conclusion with the one of Portegies Zwart (2009) in Fig. 1 we show the computed final positions of the probe particles when the spiral perturbations of the galactic gravitation field is not taken into account. Indeed, in this case after 4.6 Gyrs the stars are lined up in a ring-like segment. 
                     
\begin {figure}
\includegraphics {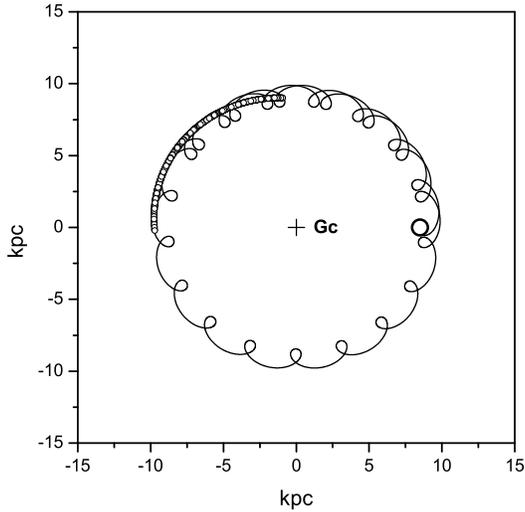}
\caption {The ring-like segment in the left-upper part of the figure produced by {\it small open circles} delineates the final positions of probe stars which started at the place shown by the {\it open large circle} at coordinates (8.5;0) (here and after the size of the open large circle is not equal to the real initial diameter of the parental cluster). Stellar motions were computed in the absence of perturbations from spiral gravitation field. The {\it cycloid-like thin solid line} demonstrates an example of epicyclic trajectory of some star from our sample. The result is shown in the coordinate system rotating at the angular rotation velocity $\Omega(r_S)$}
\label {fig1}
\end {figure}
       
Further, our results of the cluster members scatter under the influence of spiral perturbations are described. 

\subsection {Stationary spiral structure}

1)~{\it Number of spiral arms $m=2$, corotation radius $r_C = r_S$, pitch angle 
$p = - 12^{\circ}$.} Below are shown the final (4.6 Gyr after the solar cluster birth) stellar positions derived for the same initial galactocentric distance of the parental cluster ($r_{cc}=8.5$ kpc) but for 2 initial its locations relative to spiral arms. In the first case (Fig. 2), the stars happen to be spread over a very large part of the galactic disc. So, the supposition of Portegies Zwart that stars, which were born close to the Sun, can not drift far from it, in general is incorrect. Our computations demonstrate that, at the final stage, in the close vicinity of any star from this sample we can find no more than 3 - 5 siblings.
                     
\begin {figure}
\includegraphics {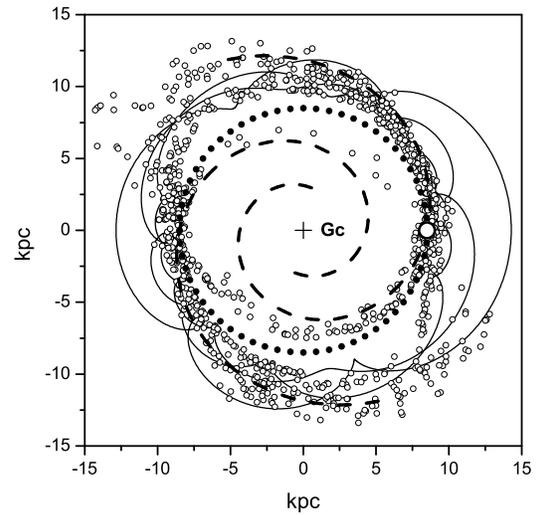}
\caption {The same as in Fig. 1, but the perturbations of the galactic gravitation field due to spiral density waves are included. The {\it ring-like line} shown by {\it filled circles} delineates the location of the corotation circle. The {\it thin solid line} is the trajectory of the same star from the sample which was shown in Fig. 1. The trajectory demonstrates significant deviation from the simple epicyclic curve. It is seen that stars are scattered over a very large part of the galactic disc. The picture is drawn in the coordinate system rotating at the angular rotation velocity $\Omega_P$.}
\label {fig2}
\end {figure}
       
In Fig. 3 we present the final positions of the cluster members for another initial azimuth angle. Here the final stellar locations on the galactic plane happen to be rather compact: there are about 70 stars in the close vicinity of which from 50 to 80 siblings can be found.
                     
\begin {figure}
\includegraphics {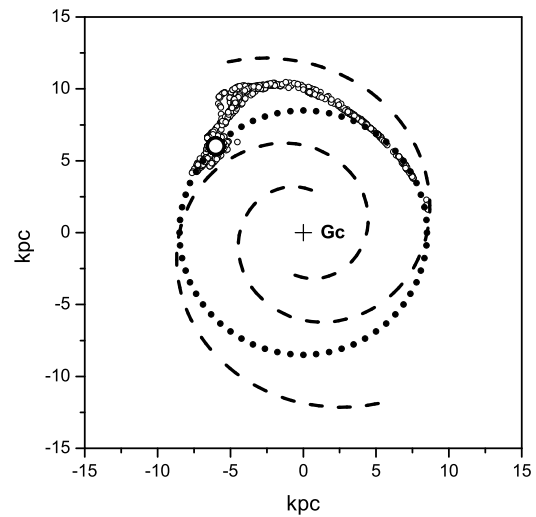}
\caption {The same as in Fig. 2, but the stars started at another galactic azimuth.}
\label {fig3}
\end {figure}

However the above space pattern of stars is destroyed if we slightly move the initial position of the cluster, say decrease the distance of its centre, $r_{cc}$,  by 0.5 kpc, keeping the same its azimuth (see Fig. 4). For this case we do not find more than 10 siblings in the close vicinity of any star from the cluster members.

\begin {figure}
\includegraphics {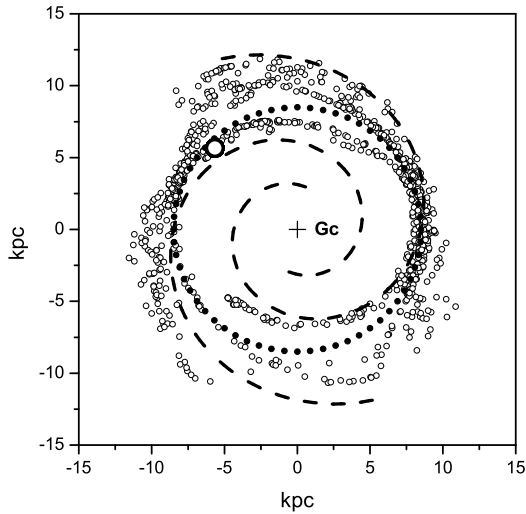}
\caption {The same as in Fig. 3, but the initial galactocentric distance of the cluster (shown by {\it large open circle}) center is 8 kpc, the initial its azimuth being the same.}
\label {fig4}
\end {figure}

2)~{\it Corotation radius $r_C = 3.5$ kpc, pitch angle $p = -12^{\circ}$.} Besides the 
spiral wave pattern with the corotation close to the present solar position, in literature one can meet models with the corotation at about 3 - 4 kpc (e.g. Weinberg 1994; Dehnen 2000, etc.). That is why we consider stellar scattering for the case with $r_C = 3.5$ kpc and 2 numbers of spiral arms.
                     
In Fig. 5 is shown the result for $m = 2$. The cluster members again happen to be scattered over a large part of the galactic disc so that in the close vicinity of any star we do not find more than 3 siblings.
                                   
\begin {figure}
\includegraphics {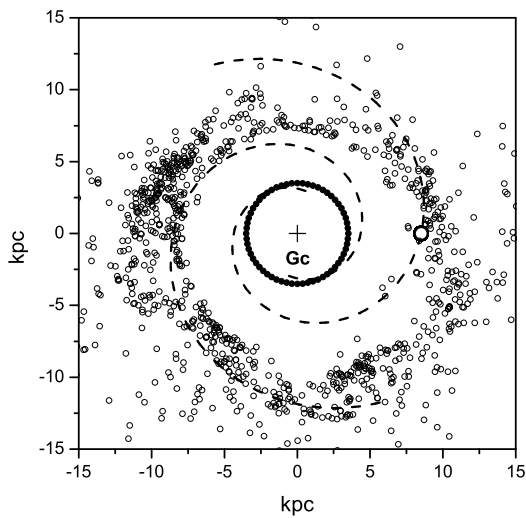}
\caption {The same as in Fig. 2, but for the corotation resonance located 
in the inner part of the galactic disc at $r_C = 3.5$ kpc 
({\it filled circles}). Number of arms $m=2$.}
\label {fig5}
\end {figure}

However for $m = 4$ at the final stage stars are assembled into a small region filling it compactly (see Fig. 6), so that for $\sim130$ stars we can reveal about 60 - 80 siblings in the close vicinity. It is obvious, such sharply distinct behavior of the stellar ensemble from the previous case is explained by more frequent oscillation (in azimuth angle) of the perturbed gravitation field which leads to less scattering effect.

\begin {figure} 
\includegraphics {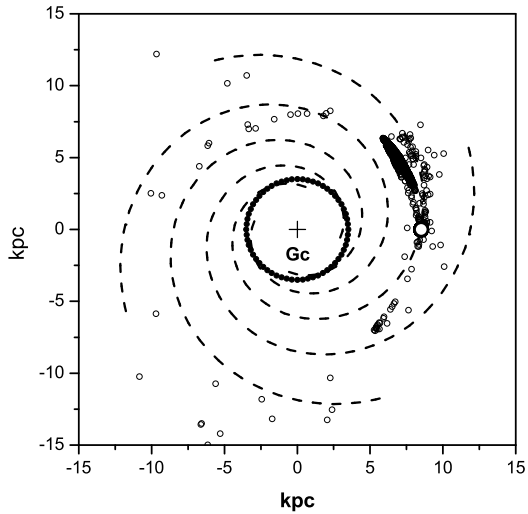}
\caption {The same as in Fig. 5, but for $m=4$.}
\label {fig6}
\end {figure}

3)~{\it Superposition of two density wave harmonics: $m = 2 + 4$ patterns.} On the basis of various observational data - Cepheid kinematics, regions of ionized hydrogen and radio emission of neutral hydrogen in 21 cm line - L\'epine et al. (2001, hereafter LMD) proposed a model of galactic spiral wave pattern as a superposition of 2 wave harmonics: $m=2$ and $m=4$ (see Fig. 3 in their paper). A similar structure was derived by Drimmel \& Spergel (2001) in their study of IR emission. Below we describe the LMD model in some details and consider the cluster members scatter.
                     
According to LMD we represent the perturbed spiral gravitation potential as:
$$\varphi_S = \varphi_2 + \varphi_4,$$
where $\varphi_m$ is the corresponding $m-th$ wave harmonic ($m=2$ or $m=4$), $\varphi_m = A_m cos(\chi_m)$, $A_m$ and $\chi_m$ are their amplitude and phase, $\chi_m = m[cot(p_m) log (r/r_0) - \vartheta +\Omega_Pt]+\Delta\chi_m$, $p_m$ is the corresponding pitch angle, $\Delta\chi_m$ is the phase shift (to set the coordinate system relative to 2-armed pattern let $\Delta\chi_2=0$, then $\Delta\chi_4=200^{\circ}$). Notice that both the patterns rotate with the same rotation velocity $\Omega_P$. We choose it so that the corotation resonance lies close to the Sun ($r_C = r_S$; as a consequence for the 4-armed pattern inner lindblad resonance is situated at about 6 kpc, the outer one at 11 kpc). In what follows, we slightly modified the parameters of LMD model since the radial scale in the last paper, $r_S$, was adopted to be 7.5 kpc whereas in the present paper $r_S = 8.5$ kpc. To adjust the pitch angle of LMD model to our present scale we assume $p_2 = -7^{\circ}$, correspondingly $p_4 = -14^{\circ}$. The wave amplitude for two-armed pattern is derived as above. The ratio of amplitudes $A_4/A_2 = 0.8$ (details see in LMD).

In the framework of this model, the final positions of the cluster members are shown in Fig. 7. Again, the stars happen to be spread over a large space volume, so in the close vicinity of any star we can reveal no more than 3 - 5 siblings. However, for another initial azimuth of the cluster we get rather compact final stellar distribution in space (Fig. 8) and among the stars one can find about 10 - 20 objects which are surrounded by 20 - 30 siblings in the vicinity of 100 pc.

\begin{figure}
\includegraphics {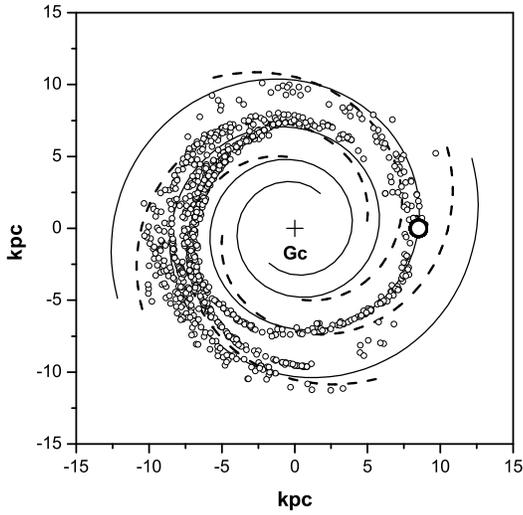}
\caption {The same as in Fig. 2, but for superposition of 2 wave harmonics $m=2+4$, see text. The spiral lines shown by {\it thin solid and dashed lines} are the loci of $min(\varphi_m)$.}
\label{fig7}
\end{figure}
                     
\begin{figure}
\includegraphics {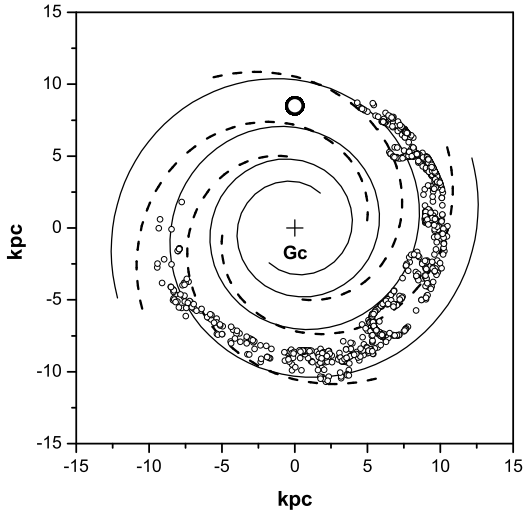}
\caption {The same as in Fig. 7, but for another initial cluster azimuth.}
\label {fig8}
\end {figure}

\subsection {Transient spiral structure}

In Fig. 9 is shown the final picture for stellar scattering in the case of "short" life-time of a pattern event of speed $\Omega_P$, i.e. $\tau_P = 100$ Myr, in Fig. 10 for "long" life-time $\tau_P = 500$ Myr. It is seen that the stars again are scattered over a very large region, so that in the close vicinity of a star we can only reveal 3 - 4 stars from the parental cluster. Our experiments show that, in general, this conclusion is kept independently of a particular set of random $\Omega_P$ and slightly depends on life-time of a pattern event.

\begin{figure}
\includegraphics {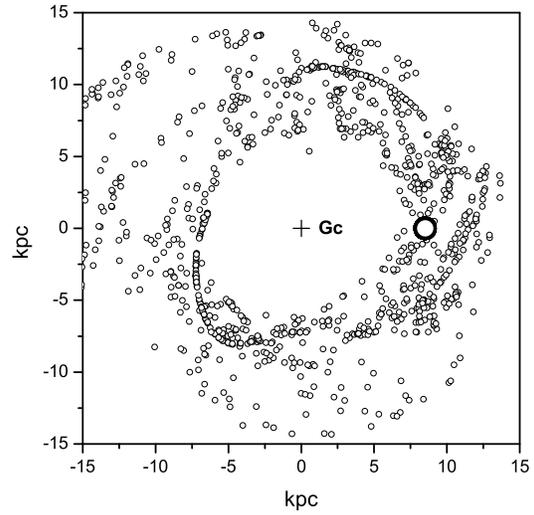}
\caption {The same as in Fig. 2, but for transient spiral structure and short life-time of a pattern with given $\Omega_P$   $\tau_P = 100$ Myr. The location of spiral arms is not shown since the pattern changes with time. The picture is shown in non-rotating coordinate system.}
\label {fig9}
\end {figure}

\begin{figure}
\includegraphics {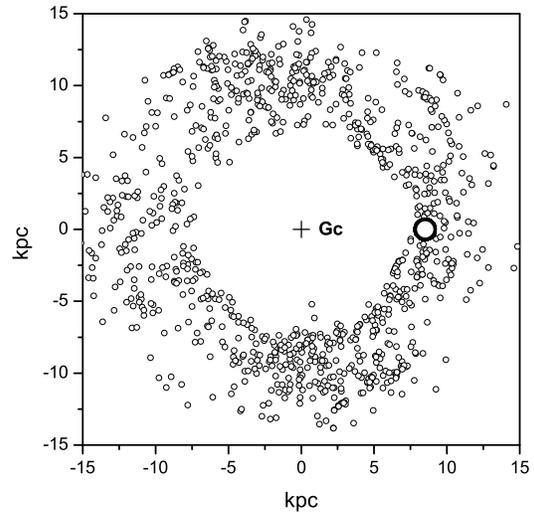}
\caption {The same as in Fig.9, but for long life-time $\tau_P = 500$ Myr.}
\label {fig10}
\end {figure}


\subsection {Open cluster decay and stellar scattering as a diffusion process}

Our results, illustrated by the above figures, show that, if we take into account the influence of spiral arms on stellar motion, in general, their trajectories are unstable: being born within small space volume, several billion years stars are scattered (predominantly chaotically) over a very large part of the galactic disc after. So we can treat the cluster stellar migration as a diffusion process and to estimate radial and azimuth diffusion coefficients. 
\footnote {Strictly speaking, except for the case with transient density waves, in our approach, stellar motion is reversible since we do not take into account their random collisions with, say, molecular clouds or other objects (transient density waves represent stochastic fluctuations of galactic gravitation fields and stellar interactions with such fluctuations are a kind of collisions). Nevertheless, random initial conditions of the parental cluster stars introduce a chaotic element in evolution of the stellar ensemble as a whole although (except for the case of transient density waves) the acting on stars force is regular. So, at least, starting with some moment of time the spreading of stars can be described in terms of diffusion despite the absence of stellar collisions.}

To derive the diffusion coefficients in radial and azimuth directions (correspondingly, $D_r$ and $D_{\vartheta}$) we consider the stellar ensemble as 
``clouds" in $r$ and $\vartheta$ spaces. The location of the cloud along $r$ or $\vartheta$ axis as a whole is determined by its mean coordinates $<r>$ and $<\vartheta>$ ($<r> =\sum r_i/N$, where $N$ is the number of probe particles in the ensemble, $r_i$ is the radial coordinate of $i - th$ star and the summation is taken over all stars from the ensemble, the similar formula should be written for mean azimuth). The typical sizes of the ``clouds", imitating the stellar ensemble in $r$ and $\vartheta$ spaces, are given by the root mean squares  $\sigma_r$ and $\sigma_{\vartheta}$, the radial dispersion as usual being $\sigma^2_r = \sum(r_i - <r>)^2/(N-1)$ and the expression for $\sigma^2_{\vartheta}$ to be written by analogy with $\sigma^2_r$. Since at the initial moment of time cloud sizes (in $r$ and $\vartheta$ directions) are small, the ones can be represented as $\delta$-functions of $r-r_{cc}$ and $\vartheta-\vartheta_{cc}$. In what follows, for simplicity we consider the diffusion process separately in $r$ and $\vartheta$ directions and use the well known one-dimensional (Cartesian - like) solution of the diffusion equation to connect the radial and azimuth dispersions with the diffusion coefficients:

$$2\sigma^2_{r, \vartheta} = \int_{0}^{t}D_{r, \vartheta}dt^{\prime}$$
(e.g. Landau \& Lifshitz 1986; their expression for the diffusion coefficient was extended for the case when the coefficient depends on time, see below).

Further, by means of our numerical experiments we compute $\sigma^2_{r, \vartheta}$ and from the above equation derive the diffusion coefficients $D_r$ and $D_{\vartheta}$.

In Figs. 11 and 12 is shown the evolution of $\sigma^2_{r, \vartheta}$ with time. It is seen that during the first 1 or 2 Gyr the both dispersions happen to be very small. Detail analysis shows that at this epoch, the dispersions vary slowly, but after 1 - 2 Gyr the above regime breaks down abruptly and $\sigma^2_{r, \vartheta}$ begin to increase sharply. So we can regard the stellar scatter as the diffusion process only for $t \ge 1-2$ Gyrs. Let us call this period of time as the {\it diffusion stage}. 

It is worthwhile to notice that the dispersions do not vary monotonically. They oscillate around some (growing) trend (the most distinctly it becomes apparent for $\sigma^2_r$).
\footnote {In case ``c" $\sigma^2_r$ stepwise increases from very low value to $\sim1$ kpc$^2$ after about 1 Gyr and then oscillates, the trend being almost flat. This means that in radial direction the size of the ensemble does not grow in average, it simply oscillates at some new mean value ($\vartheta$-dispersion demonstrates the growing trend but it is dozens of times less than in other cases, see Fig. 12). This result reflects the more or less compact final positions of the probe stars in Fig. 8.}

The above pictures suggest that the behavior of our system should be understood in terms of dynamical chaos. However, the detail analysis of this question is beyond the scope of our paper.

Approximation of the time dependence of $\sigma^2_r$ at the diffusion stage by the linear trend corresponds to constant $D_r$ (in kpc$^2$ Gyr$^{-1}$): $\sim$~1.6 (case ``a"); $\sim$~1.4 (case ``b") and $\sim$~4.2 (case ``d"). 
\footnote {Case ``c" we exclude from interpretation in terms of diffusion.}
As it was expected, the largest stellar migration takes place in the variant of transient density waves: the corresponding diffusion coefficient more than 2 times exceeds $D_r$ derived in other models. By means of $D_r$ we can estimate a region $\Delta r$ which will be occupied by stars, born in an open cluster, during the life-time of the galactic disc $T_d\sim10$ Gyrs. The region is: $\Delta r \sim (D_rT_d)^{1/2}\sim 4 - 6$ kpc.

The diffusion stage in azimuth evolution of the stellar ``cloud"  comes after  $\sim$2 Gyrs (see Fig. 12; for the case ``c" $\sigma^2_{\vartheta}$ occurs to be about 20 times less than in other cases, so we will not interpret this case in terms of diffusion although the growing trend is obvious). However, unlike $\sigma^2_r$ the azimuth dispersion $\sigma^2_{\vartheta}$ grows in time with acceleration, at least, quadratically, so that $D_{\vartheta}$ (in rad$^2$ Gyr$^{-1}$) happens to be: $\approx -22+11t$ (case ``a"); $\approx -23+9t$ (case ``b"); and $\approx -15+4t$ (case ``d"). In azimuth direction the rms effective cloud radius sometimes exceeds 360 $^{\circ}$  which means that many stars overtake others more than one period.

It is interesting to explore in more detail the evolution of the dispersions at the very early epoch. Indeed, according to Bland-Hawthorn et al. (2010), the most part of forming open clusters are unbound and analysis of various data shows that most of them are disrupted during first 100 Myr. By means of our experiments we can estimate the decay time $\Delta t_{r,\vartheta}$ during which the typical cluster size ($2\sigma_{r,\vartheta}$) increases, say 10 times (this threshold value we interpret as a border which separates the state when the stars still belong to a cluster, from the one when they have become a part of the galactic disc.).

There is an important difference in time evolution of $r$ and $\vartheta$ dispersions. From Figs. 11,12 it is seen that at the diffusion stage both the dispersions oscillate with time. In general, such behavior is kept at early epoch, but during several first tens Myr after the open cluster formation $\sigma^2_{\vartheta}$ grows almost monotonically. So, we can state: for cases ``a, b, d" the azimuth decay time $\Delta t_{\vartheta}$ is of the order of 50 - 70 Myr whereas for the radial one $\Delta t_r$ happens to be $\sim$1.1 - 1.4 Gyr. Such large difference in the corresponding typical time periods is obviously connected with different radial and azimuth resilience of stellar ensemble as a part of the galactic disc.

\begin{figure}
\includegraphics {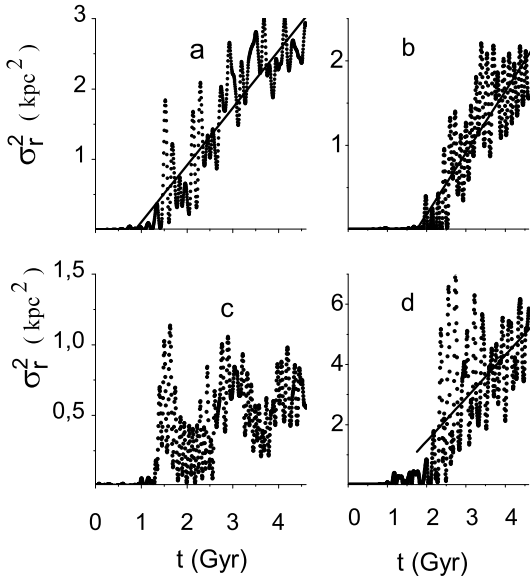}
\caption {{\it Filled circles}: evolution in time the radial dispersion $\sigma^2_r$. Case "a" corresponds to Fig. 2; "b" - to Fig. 4; "c" - to Fig. 8; "d" - to Fig. 9.}
\label {fig11}
\end {figure}

\begin{figure}
\includegraphics {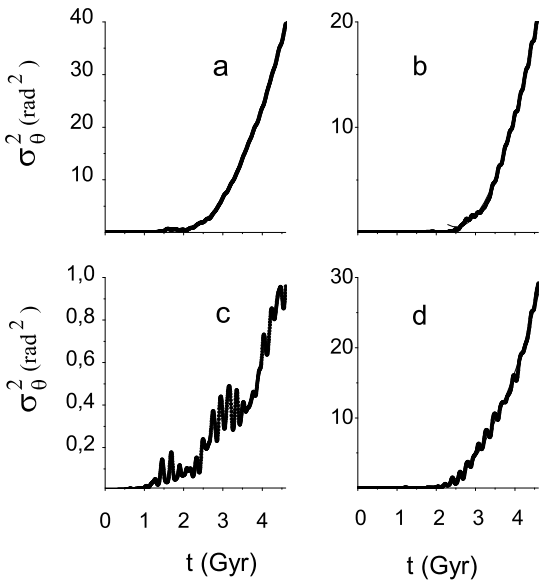}
\caption {The same as in Fig. 11 but for $\sigma^2_{\vartheta}$.}
\label {fig12}
\end {figure}


       \section {Conclusion}

The problem of searching solar siblings is a very exciting but difficult one. Indeed, we have to recognize the members of solar family among the close $\sim10^5$ stars. The only chance is to find a group of, say, several tens stars with particular properties: close ages and chemical element abundances, special velocity pattern in space, etc.

In the present paper, we tried to answer the question: is the situation so optimistic as it was inferred by Portegies Zwart (2009)? Our conclusion is as follows. If we take into account perturbations of the galactic gravitation field from spiral density waves, stellar motion strongly deviates from quasi-circular epicyclic trajectory. In general, Portegies Zwart's supposition that stars do not drift far from the Sun, is broken. Unlike Portegies Zwart's opinion, stars that were born within a small space volume can drift far from the cluster members. In spite of that they have small differences in their initial positions ($\le6$ pc) and the same initial velocities, in general, after 4.6 Gyrs the siblings are scattered over an unexpectedly large part of the galactic disc. 
Most of our experiments demonstrate that in the close vicinity of the Sun we do not have a hope to discover more than several solar relatives if we adopt that the solar parental cluster had $\sim 10^3$ stars as in Portegies Zwart (2009) model. Notice also that we did not take into account the effects of additional stellar scattering due to interaction with molecular clouds and velocity dispersion.

Nevertheless, in some cases we derive sufficiently compact final locations of siblings, so that in the close vicinity of the Sun we have a hope to reveal several tens of the parental cluster members. 

However, if we adopt that the parental cluster had $\sim~10^4$ stars (instead of $10^3$, like in Portegies Zwart model) the task becomes less hopeless.

\section {Acknowledgments}

Authors are grateful to anonymous referee for very important comments.

The paper was supported by the grant No 02.740.11.0247 of Federal agency for science and innovation of the Russian Ministry of Education.
                    

\label{lastpage}
       
\end{document}